# ThermSense: Smartphone-based Breathing Sensing Platform using Noncontact Low-Cost Thermal Camera

Youngjun Cho, Nadia Bianchi-Berthouze, Simon J. Julier, Nicolai Marquardt
*University College London*
*London, United Kingdom*
{youngjun.cho.15, nadia.berthouze, s.julier, n.marquardt}@ucl.ac.uk

*Abstract*—The ability of sensing breathing is becoming an increasingly important function for technology that aims at supporting both psychological and physical wellbeing. We demonstrate ThermSense, a new breathing sensing platform based on smartphone technology and low-cost thermal camera, which allows a user to measure his/her breathing pattern in a contact-free manner. With the designed key functions of *Thermal Voxel Integration*-based breathing estimation and respiration variability spectrogram (RVS, bi-dimensional representation of breathing dynamics), the developed platform provides scalability and flexibility for gathering respiratory physiological measurements ubiquitously. The functionality could be used for a variety of applications from stress monitoring to respiration training.

*Key words*—Noncontact Biosensor; Breathing Sensor; Contact-free Physiology Measurement; Thermal Camera.

## 1. Introduction

Non-contact thermal imaging computes dynamic temperature reflecting the radiated power of electromagnetic waves emitted from an object [1]. Thanks to advancements in thermal imaging technologies, thermal cameras have been employed in a variety of sectors, such as physiology measurements (e.g., respiration [2], [3], skin conductance [4]) and affective computing (e.g., mental stress recognition [5], arousal-valence detection [6]). In comparison with other types of contact-less physiology sensors (e.g., remote photoplethysmography (rPPG) [7]), thermal imaging has less constraints (e.g., no illumination or privacy issues). In addition, small, light weight and cheap characteristics of recently launched low-cost thermal cameras (e.g., FLIR One, Seek Thermal) have tackled its known barriers (e.g., heavy weighted, expensive) for real-world situations. A recent body of work [2], [5], [8] also support the use of low-cost thermal imaging by enhancing the low signal quality of the cameras. Building on this work, we propose *ThermeSense*, a low-cost smartphone-based thermal imaging breathing sensing platform.

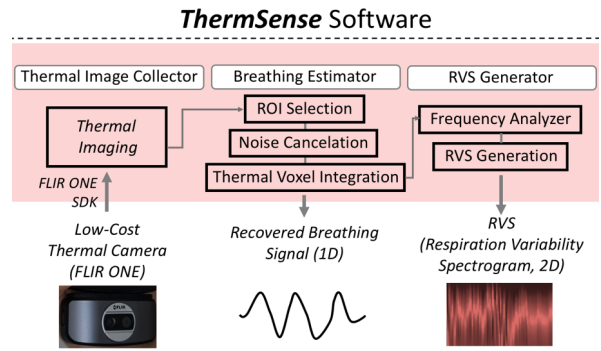

Figure 1. Software architecture of ThermSense: it cosists of three parts – continuous thermal image extractor, breathing estimator and respiration variability spectrogram (RVS) generator.

## 2. System Architecture

Figure 1 shows the overall software architecture of ThermSense. It mainly consists of *thermal image extractor*, *breathing estimator* and *respiration variability spectrogram (RVS) generator*. Among low-cost thermal cameras commercially available, FLIR One (for Android - dimensions: 72mm x 26mm x 18mm, FLIR), which is one of the cheapest devices in 2017, was chosen for this platform. This thermal camera can be used as a smartphone accessory – thermal image processing can be run on a smartphone in real time. It detects electromagnetic waves in the spectral range of 8 to 14μm with a spatial resolution of 160x120 and a temporal resolution of less than 9 frames per second (fps). The *thermal image extractor* extracts raw absolute temperature information through the use of functions built on its sdk[1]. Here, the emissivity is set to 0.98 that is the emissivity value for the human skin [9]. The *breathing estimator* analyses the thermal distribution of each frame to recover the one-dimensional breathing signal. Lastly, the recovered signal is feed-forwarded to the *RVS generator*. This is to produce bi-dimensional Respiration Variability Spectrogram signatures (proposed in [5]) condensing a person's breathing dynamic. More details about the breathing estimation and RVS are explained in the next sections.

---

[1] *http://developer.flir.com/flir-one-software-development-kit/*

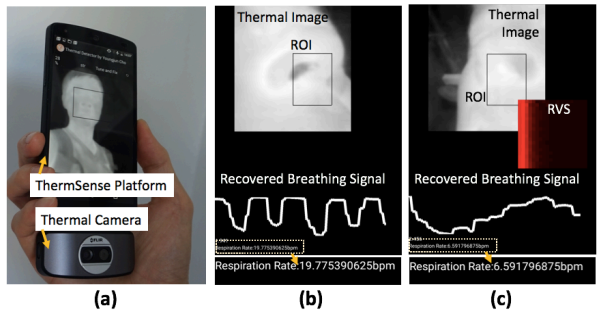

Figure 2. (a) ThermSense platform running on an Android smartphone,(b),(c)ThermSense screenshots: thermal image is produced reflecting the human emissivity, raw breathing signals are recovered and shown in the bottom, breathing rates are displayed (30 seconds) and respiration variability spectrograms (RVSs) are generated from a series of real time processing.

The system has two key functions detailed below:

***Estimation of Breathing Pattern:*** An example of physical interfaces for supporting ThermSense is shown in Figure 2(a). Using this type of setups, a user can measure, and see, his breathing pattern in real time. Several computational stages support this process. The first stage is the ROI selection. In this version of the platform, this selection is manually performed to minimise computational resources, although some techniques (i.e., automated ROI tracking) can enable this process to be done automatically [2], [3]. Given the selected ROI, noise cancelation and Thermal Voxel Integration [2] stages are run in a row to produce one-dimensional breathing signals. These two stages are based on techniques proposed in [2]. Signal processing filters (e.g., digital bandpass filter) directly support to cancel noises in thermal distribution sequences. Thermal Voxel Integration is a state-of-the-art method for recovering one's breathing pattern by treating a unit temperature as a three-dimensional voxel and integrating the thermal voxels along with the breathing cycle [2]. For technical details we refer to [2]. A series of this procedure in turn produce one-dimensional time-varying breathing signals (Figure 2(b) and 2(c)).

**Generation of Respiration Variability Spectrogram:** Respiration Variability Spectrogram (RVS) is a recently proposed bi-dimensional representation of breathing dynamics in [5]. This form of spectrogram technically condenses breathing dynamic information and updates itself by analysing the recovered one-dimensional breathing signal. An example RVS is shown in Figure 2(c). This can be directly inputted to convolutional neural networks based feature learning models to extract meaningful features in relation to affective states [5]. For the implementation and evaluation studies see [2], [5].

## 3. Application: Mental Stress Detection

ThermSense can support a variety of real life applications. Mental and physical stress monitoring is a key issue targeted by the affective computing community in many contexts: health, wellbeing and entertainment. Our platform allows to monitor such states by reducing outdoor and night illumination constraints (i.e., key limitations of remote PPG [7]) and removing the need to wear sensors (e.g., PPG wrist band [10]) that may not be suitable or are unpleasant. For example, in the gym, people may find sensors unpleasant due to sweat or constrained movements. With feedback technologies available (e.g., tactile and sound feedback [11], [12]), ThermSense can be useful for enhancing a person's stress awareness. Other contexts are: sleeping infants at night (the camera attached to the cradle), people with a certain type of pain (e.g., CRPS [13]) tend to reduce the number of clothes and objects that touch their skin. Based on its simple and easy set ups, ThermSense could drive the scalability and reliability of thermal imaging-based breathing sensing in real world settings.

## Acknowledgments

Youngjun Cho was supported by University College London Overseas Research Scholarship (UCL-ORS) awarded to top quality international postgraduate students.